\documentclass[pra,bibnotes,aps,twocolumn,longbibliography,superscriptaddress,nofootinbib]{revtex4-1}

\usepackage{amsmath,amsfonts,amssymb,amsthm,mathtools}
\usepackage[locale=US]{siunitx}
\usepackage{graphicx}
\usepackage{color}
\usepackage[colorlinks]{hyperref}
\usepackage{bm}
\usepackage{bbold}
\usepackage{float}
\usepackage{placeins}
\usepackage[abs]{overpic}
\usepackage{textcomp}

\newcommand{\hmf}{h_{\mathrm{mf}}}
\newcommand{\opt}{\mathrm{opt}}
\newcommand{\rf}{\mathrm{rf}}
\newcommand{\mf}{\rm{mf}}
\newcommand{\annd}{\hat a^\dagger}
\newcommand{\ann}{\hat a}

\newtheorem{theorem}{Theorem}

\DeclareMathOperator{\Tr}{Tr}

\begin{document}
	
\title{Excited-state quantum phase transitions in spinor Bose-Einstein condensates}
	
\author{Polina Feldmann}
\email{polina.feldmann@itp.uni-hannover.de}
\affiliation{Institut f\"ur Theoretische Physik, Leibniz Universit\"at Hannover, Appelstr. 2, 30167 Hannover, Germany}
\author{Carsten Klempt}
\affiliation{Institut f\"ur Quantenoptik, Leibniz Universit\"at Hannover, Welfengarten 1, 30167 Hannover, Germany}
\author{Augusto Smerzi}
\affiliation{QSTAR, INO-CNR, and LENS, Largo Enrico Fermi 2, 50125 Firenze, Italy}
\author{Luis Santos}
\affiliation{Institut f\"ur Theoretische Physik, Leibniz Universit\"at Hannover, Appelstr. 2, 30167 Hannover, Germany}
\author{Manuel Gessner}
\affiliation{Laboratoire Kastler Brossel, ENS-Universit\'e PSL, CNRS, Sorbonne Universit\'e, Coll\`ege de France, 24 Rue Lhomond, 75005 Paris, France}
	
\begin{abstract}
    Excited-state quantum phase transitions (ESQPTs) extend the notion of quantum phase transitions beyond the ground state. They are characterized by closing energy gaps amid the spectrum. Identifying order parameters for ESQPTs poses however a major challenge. We introduce spinor Bose-Einstein condensates as a versatile platform for studies of ESQPTs. Based on the mean-field dynamics, we define a topological order parameter that distinguishes between excited-state phases, and discuss how to interferometrically access the order parameter in current experiments. Our work opens the way for the experimental characterization of excited-state quantum phases in atomic many-body systems.
\end{abstract}

\maketitle
\date{\today}

Quantum phase transitions~(QPTs) are sudden changes in the ground-state properties of a system. The
ground-state energy and wave function behave non-analytically and the gap between the ground state and the first excited state closes when, at zero temperature, a control parameter is adiabatically varied across a critical value~\cite{Sachdev2011}. The idea of QPTs has been extended in recent years to out-of-equilibrium quantum many-body systems~\cite{PhysRevLett.105.015702,Heyl2018}. For example, a sudden shift of a parameter~(quantum quench), can lead to dynamical QPTs, which are characterized by a non-analyticity of physical quantities as a function of time~\cite{Heyl2018}. A direct generalization of QPTs beyond the ground state is given by excited-state quantum phase transitions~(ESQPTs)~\cite{Cejnar2006, Iachello2008, Cejnar2020}. Their distinguishing signature is a closing gap at nonzero energies: excited states cluster at a critical energy, which leads to a singularity in the density of states (DOS). Typically, the critical energy is a continuous function of a control parameter.
Thus, in contrast to ground-state QPTs, ESQPTs can be crossed both by varying a control parameter at constant energy and by 
varying the energy at fixed parameters.

ESQPTs have been theoretically studied in a large variety of many-body quantum systems~\cite{Iachello2008, Stransky2014, Stransky2015}, including the Lipkin-Meshkov-Glick (LMG) model~\cite{Leyvraz2005}, Dicke and Jaynes-Cummings models~\cite{Perez-Fernandez2011, Brandes2013}, interacting boson models~\cite{Cejnar2006, Iachello2008, Macek2019}, molecular bending transitions~\cite{PerezBernal2008, Larese2011}, and the quasi-energy spectrum of driven systems~\cite{Bastidas2014}. Experimentally, ESQPTs have been confirmed in microwave Dirac billiards~\cite{Dietz2013} and in molecular spectroscopy~\cite{Zobov2005,PhysRevLett.95.243002}. Signatures of ESQPTs have been predicted in the many-body dynamics after a quench~\cite{Perez-Fernandez2011, Santos2015, Kloc2018} and in time-averaged expectation values~\cite{PhysRevA.91.013631}. 
However, identifying order parameters that distinguish neighboring excited-state quantum phases from each other remains a challenge~\cite{Cejnar2006, Iachello2008}.

Spinor Bose-Einstein condensates~(BECs) attract since several years a major interest as an exceptional tool for the study of many-body quantum dynamics~\cite{Ueda2012,RevModPhys.85.1191}, including coherent spinor dynamics~\cite{Chang2005}, classical bifurcations~\cite{Zibold2010}, and the generation of highly-entangled many-body states~\cite{Luecke2011, Gross2011, Luo2017, Pezze2019}. So far systematic investigations of critical behavior in such systems have focused on the ground-state QPTs~\cite{Ueda2012,PhysRevLett.102.125301,PhysRevLett.107.195306,PhysRevLett.111.180401,Luo2017}, though observations of diverging oscillation periods can be interpreted as signatures of ESQPTs~\cite{You2005,Zhao2014}. Very recently, a study of the quench dynamics of a spinor BEC revealed a dynamical QPT, which has been related to a phase transition in the highest-energy level~\cite{Tian2020}. 

In this Letter, we propose spinor BECs as a platform to explore ESQPTs in a paradigmatic class of models. We identify ESQPTs in a ferromagnetic spin-1 BEC, and show that the different excited-state quantum phases can be distinguished by the topology of classical phase-space trajectories. We use this to introduce an order parameter that is related to the dynamics of coherent states. This order parameter can be accessed by interferometry in existing experimental setups. 
Our work is, hence, an important step towards the characterization of excited-state quantum phases and towards the systematic exploration of ESQPTs with controllable many-body quantum systems.


\begin{figure*}
	\begin{flushleft}
	\begin{minipage}[b]{0.497\textwidth}
	\begin{center}
		\begin{overpic}[scale=0.52]{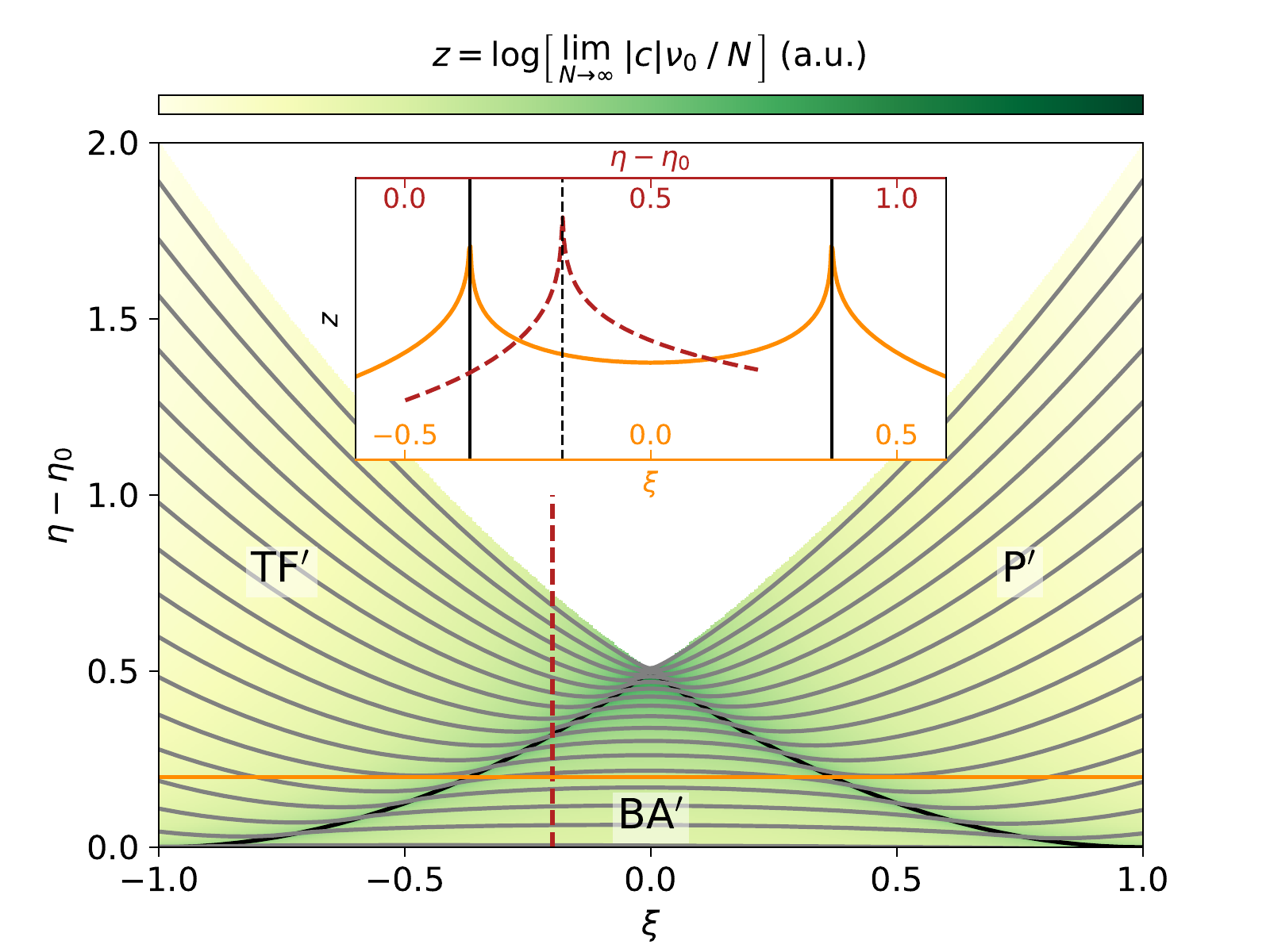}
			\put (0,159) {(a)}
		\end{overpic}
	\end{center}
	\end{minipage}	
	\hspace{4em}
	\begin{minipage}[b]{0.2\textwidth}
		\begin{overpic}[scale=0.32]{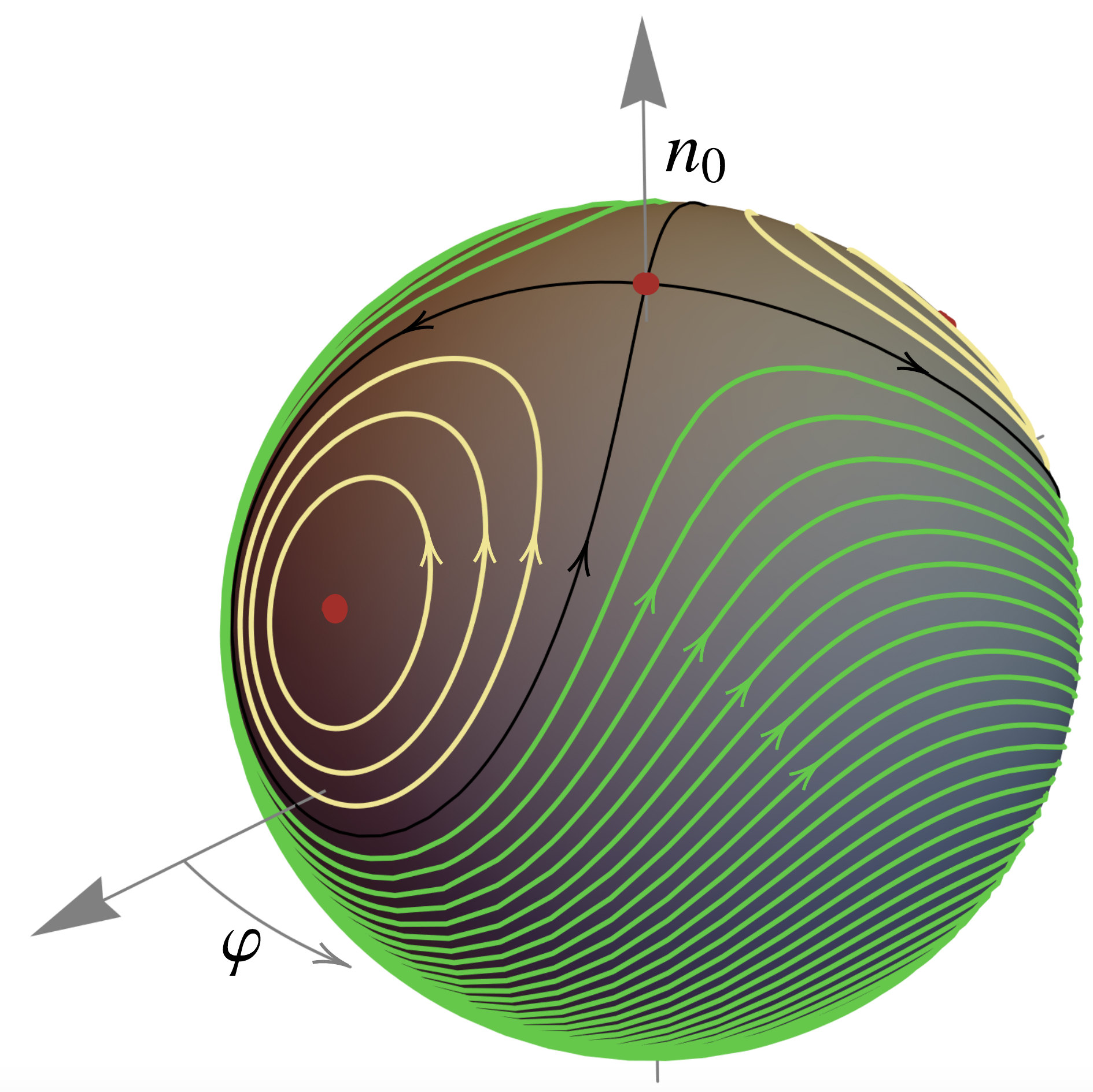}
			\put (4,159) {(b)}
		\end{overpic}
	\end{minipage}
	\end{flushleft}		
	\caption{Excited state quantum phases of a ferromagnetic spin-1 BEC with zero magnetization. (a) DOS in the mean-field limit as a function of $\xi$ and $\eta-\eta_0$ with $\eta_0(\xi)=-\frac{1}{2}(\xi^2+1)$. The ESQPTs at $\eta_*=-|\xi|$ (black) divide the $\xi$-$\eta$-plane into three phases: the TF$'$ phase, the P$'$ phase, and the BA$'$ phase. The DOS diverges at the ESQPTs. The inset shows the DOS along lines of constant $\xi=-0.2$ (red, dashed) and $\eta-\eta_0=0.2$ (orange, solid). The spectrum of a BEC of $N=100$ atoms (gray, every third eigenvalue) exhibits avoided crossings at the ESQPTs. (b) Classical phase space and trajectories for $\xi=0.5$. The separatrix (black) separates trajectories in the P$'$ phase ($\eta>\eta_*$, green) with winding number $w=1$ from trajectories in the BA$'$ phase ($\eta<\eta_*$, yellow) with $w=0$. Stationary points of $\hmf$ are marked in red.}
	\label{fig:1}
\end{figure*}
	

{\it Ground-state quantum phases.---}We consider a ferromagnetic spin-$1$ BEC of $N$ atoms with three spin states $m=\pm 1,0$. We assume a tight enough external trapping of the BEC such that, to a good approximation, all spin states share a common spatial mode~(single-mode approximation). The spin degrees of freedom are then well described by the Hamiltonian density~\cite{Ueda2012}
\begin{align}
    \label{eq:HamiltonianQ}
    \hat h &= \frac{q}{2N}\left(N-2\hat N_0\right)\\
    &+\frac{c}{N^2}\left[
    \hat a_1^\dagger \hat a_{-1}^\dagger \hat a_0^2+\hat a_0^{\dagger 2}\hat a_1\hat a_{-1}
    +\hat N_0\!\left(\!N\! -\! \hat N_0+\frac{1}{2}\!\right)\! +\! \frac{\hat D^2}{2}\right]\!,\notag
\end{align}
where  $\hat a_m^\dagger$ and $\hat a_m$ are the bosonic creation and annihilation operators for state $m$, $\hat N_m\equiv \hat a^\dagger_m \hat a_m$ with $\sum_m \hat N_m=N$, and $\hat D\equiv \hat N_1 - \hat N_{-1}$ is the magnetization. 
The interaction strength $c$ depends on the spatial wave function and on the mass and scattering lengths of the atoms. A ferromagnetic BEC is characterized by $c<0$~\cite{Ueda2012}. 
The effective quadratic Zeeman shift $q$ incorporates microwave dressing and thus may be both positive and negative~\cite{Zhao2014}.
The linear Zeeman effect has been eliminated by moving to a rotating frame.   
The Hamiltonian density~$\eqref{eq:HamiltonianQ}$ conserves $\hat D$ and the parity $\hat I=(-1)^{\hat N_0}$. In the eigenspace of $\hat D$ with eigenvalue $D=0$, $\hat h$ features three ground-state phases~\cite{Ueda2012,Luo2017} depending on the ratio $\xi\equiv \frac{q}{2|c|}$: the Twin-Fock~(TF) phase for $\xi<-1$, the Polar~(P) phase for $\xi>1$, and the Broken-Axisymmetry~(BA) phase for $|\xi|<1$. 	


{\it Excited-state quantum phases.---}To reveal the excited-state phases, we study the mean-field limit~\cite{WernerRaggio89, WernerDuffield92a, WernerDuffield92b, appendix} $N\to\infty$ of Model~\eqref{eq:HamiltonianQ} for the case of zero magnetization.
We introduce the coherent states $|\bm{\alpha},N\rangle\equiv \frac{1}{\sqrt{N!}}(\sum_m\alpha_m\hat a_m^\dagger)^N|0\rangle$, where $\bm\alpha\equiv(\alpha_{1},\alpha_0,\alpha_{-1})$, $\alpha_m\equiv\sqrt{n_m}\operatorname{e}^{i\phi_m}$, $n_m\geq 0$, $\phi_m\in[0,2\pi)$, and $\sum_m\!n_m=1$. The coherent states with $\langle\hat D\rangle=0$, i.\,e., $|\alpha_1|^2=|\alpha_{-1}|^2$, yield the classical Hamiltonian~\cite{appendix}
\begin{align}
    \begin{split}
    \label{eq:HamiltonianC}
    \frac{\hmf(\bm{\alpha})}{|c|}& = \frac{1}{|c|}\lim_{N\to\infty}\langle\bm{\alpha},N|\hat h|\bm{\alpha},N\rangle\\&=\xi(1-2n_0)-2n_0(1-n_0)\cos^2\!\phi,
    \end{split}
\end{align}
where $\phi\equiv \phi_0- (\phi_1 + \phi_{-1})/2$. Note that parity conservation results in $\hmf(\phi+\pi)=\hmf(\phi)$.	
The mean-field dynamics is governed by the equations of motion~\cite{Ueda2012, You2005, appendix}
\begin{align}
    \begin{split}
    \label{eq:EOMexplicit}
    &\frac{\operatorname{d}}{\operatorname{d}\!\tau}n_0=\frac{\partial}{\partial\phi}\frac{\hmf}{|c|},\;\; \frac{\operatorname{d}}{\operatorname{d}\!\tau}\phi=-\frac{\partial}{\partial n_0}\frac{\hmf}{|c|},\;\;\text{and}\\& \frac{\operatorname{d}}{\operatorname{d}\!\tau}(\phi_1-\phi_{-1})=0
    \end{split}
\end{align}  
with $\tau\equiv |c|t/\hbar$. The mean-field limit of the DOS $\nu_0(\eta)$ in the $D=0$ subspace can be computed according to~\cite{appendix}
\begin{align}
    \begin{split}
    \lim\limits_{N\to\infty}|c|\frac{\nu_0(\eta)}{N}=\int\!\!\mathcal{D}\bm{\alpha}\delta\!\left(n_1-n_{-1}\right)\,\delta\!\left(\!\frac{\hmf(\bm{\alpha})}{|c|}-\eta\!\right),
    \label{eq:DOSReduced}
    \end{split}
\end{align}	
where $\mathcal{D}\bm{\alpha}\equiv \frac{1}{(2\pi)^3}\prod_m\!\operatorname{d}\!n_m\!\operatorname{d}\!\phi_m\,\delta\!\left(\sum_m\!n_m-1\right)$ and $\eta$ denotes the energy divided by $N|c|$.
Below we employ Eqs.~\eqref{eq:EOMexplicit} and~\eqref{eq:DOSReduced} to study the signatures of ESQPTs.

Extending the ground-state phase diagram to the entire energy spectrum, we identify three excited-state phases in the $\xi$-$\eta$-plane: the TF$'$ phase for $\eta > -|\xi|$ and $\xi<0$, the P$'$ phase for $\eta>-|\xi|$ and $\xi>0$, and the BA$'$ phase for $\eta<-|\xi|$. The phases are indicated in Fig.~\ref{fig:1}a, where we have subtracted $\eta_0(\xi)=-\frac{1}{2}(\xi^2+1)$, which corresponds to the ground-state energy in the mean-field limit, from $\eta$. The excited-state phases are separated by ESQPTs at $\eta_*=-|\xi|$ with $0<|\xi|<1$. In the limit $|\xi|\to 0$, $\eta_*$ hits the maximum of $\hmf/|c|$. As $|\xi|$ approaches $1$, the ESQPTs evolve into the known ground-state QPTs. 


{\it Signatures of ESQPTs.---}As expected for ESQPTs~\cite{Cejnar2006,Iachello2008}, the DOS~\eqref{eq:DOSReduced} diverges at $\eta_*(\xi)$. Fig.~\ref{fig:1}a displays the mean-field DOS as a function of $\xi$ and $\eta-\eta_0$. Furthermore, it shows that in a finite-size system the ESQPTs reveal themselves by a sequence of avoided crossings in the energy spectrum~\cite{Cejnar2006}. The divergence of the DOS is due to stationary points of $\hmf$. At a stationary point, $\frac{\partial}{\partial\phi}\hmf=\frac{\partial}{\partial n_0}\hmf=0$ causes the integrand in Eq.~\eqref{eq:DOSReduced} to become singular. There are three stationary points at each $0<|\xi|<1$: a saddle point at $\eta_*$ and two minima at $\eta_0$. The saddle point is located at $n_0=0$ for $\xi<0$ or at $n_0=1$ for $\xi>0$, and the minima are at $n_0=(\xi+1)/2$ and $\cos^2(\phi)=1$, see Fig.~\ref{fig:1}b. Note that these stationary points do not depend on the restriction to coherent states with $\langle\hat D\rangle=0$. However, the unrestricted DOS~\cite{appendix} remains finite at $\eta_*$. 

The phase-space trajectories~\cite{appendix} of $\hmf$ provide further signatures of the ESQPTs. The classical phase space is a sphere with $z$-axis $n_0$ and azimuthal angle $\phi$. Figure~\ref{fig:1}b shows exemplary trajectories for $\xi=0.5$. The trajectories reflect the symmetry $\hmf(\phi+\pi)=\hmf(\phi)$.
Since $\hmf(\xi,n_0,\phi)=\hmf(-\xi,1-n_0,\phi)$, for $\xi<0$ the phase space would appear upside down. As in the LMG model~\cite{Ribeiro2008}, the sets of trajectories at fixed $\xi$ and $\eta$ (the energy hypersurfaces) change topology at $\eta_*(\xi)$---at the critical energy hypersurfaces called separatrices. For $\eta>\eta_*$, i.\,e., in the TF$'$ and P$'$ phases, there is only one trajectory per $\xi$ and $\eta$. By contrast, for $\eta<\eta_*$, i.\,e., in the BA$'$ phase, the evolution can follow one of two disconnected trajectories. Each of these trajectories breaks the classical symmetry $\hmf(\phi+\pi)=\hmf(\phi)$. Note, however, that the corresponding quantum symmetry $I$ cannot be broken in the $D=0$ subspace, where all states belong to a single eigenspace of $I$. 


\begin{figure*}
    \begin{minipage}{0.497\textwidth}
    \begin{center}
    \begin{overpic}[scale=0.52]{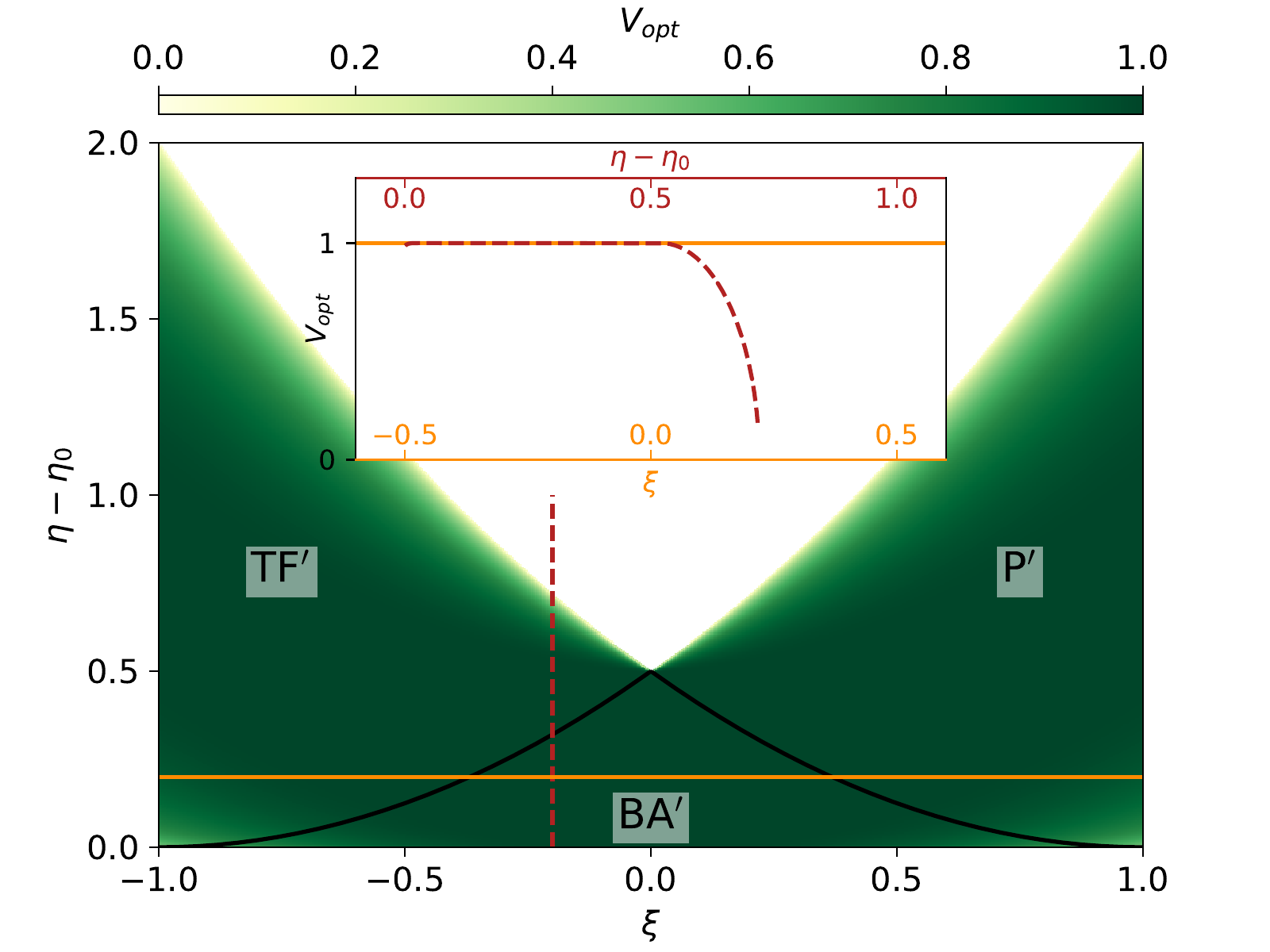}
    \put (0,159) {(a)}
    \end{overpic}
    \end{center}
    \end{minipage}
    \begin{minipage}{0.497\textwidth}
    \hspace*{0.5em}
    \begin{overpic}[scale=0.52]{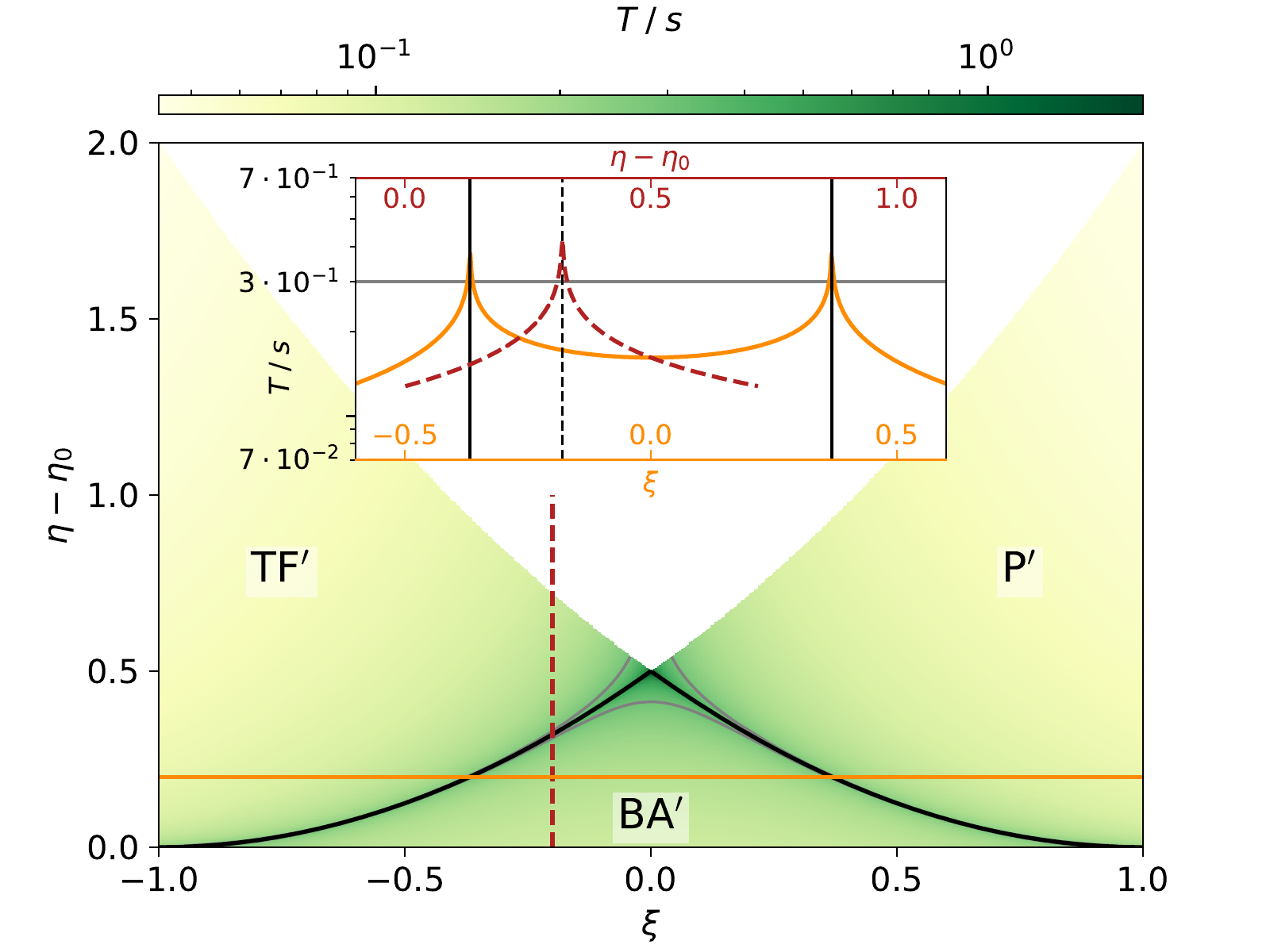}
    \put (0,159) {(b)}
    \end{overpic}
    \end{minipage}
    \caption{Measuring $p=\cos(\pi w)$ to distinguish adjacent excited-state quantum phases requires a large optimized visibility $V_{\opt}$ and a short periodicity $T$. (a)~$V_{\opt}$ is large throughout the vast majority of the phase diagram. (b)~$T$ for $|c|/\hbar=2\pi\times\SI{4}{Hz}$. A moderate value of \SI{0.3}{s} (gray) is surpassed only at the immediate vicinity of the ESQPTs. (a, b)~Black lines mark the ESQPTs. The insets show $V_{\opt}$ and $T$ along lines of constant $\xi=-0.2$ (red, dashed) and $\eta-\eta_0=0.2$ (orange, solid).}
    \label{fig:2}
\end{figure*}


{\it Order parameter.---}The solutions  $n_0(t)$ and $\phi(t)$ of the classical equations of motion, Eq.~\eqref{eq:EOMexplicit}, are periodic~\cite{Ueda2012, You2005, appendix}. In the TF$'$ and P$'$ phases, the phase-space trajectories encircle the $n_0$-axis~(green curves in Fig.~\ref{fig:1}b)---clockwise in the TF$'$ phase and counterclockwise in the P$'$ phase. By contrast, the trajectories in the BA$'$ phase do not enclose the $n_0$-axis~(yellow curves). We define our order parameter $w$ as the winding number of the classical trajectories with respect to the $n_0$-axis, such that $w=-1$ in the TF$'$, $w=1$ in the P$'$, and $w=0$ in the BA$'$ phase. We observe that $w$ can be expressed in a particularly simple form. Let us denote the period of $n_0(t)$ at fixed $\xi$ and $\eta$ by $T$. In the BA$'$ phase, the periods of $\phi(t)$ and $n_0(t)$ coincide and, thus, $\phi(t+T)=\phi(t)$. In the TF$'$ and P$'$ phases, however, $\phi(t+T)=\phi(t)\pm \pi$. Hence, 
\begin{align}
    \begin{split}
    w=\frac{1}{\pi}[\phi(T)-\phi(0)].
    \end{split}
\end{align}
In contrast to most observables that have been studied in the context of ESQPTs~\cite{Heiss2005, Ribeiro2008, Brandes2013, Iachello2008, Perez-Fernandez2011, Bastidas2014, Santos2015, Kloc2018}, $w$ is not merely singular at the phase transitions. It qualitatively distinguishes the entire excited-state phases by the dynamics of coherent states.

In the following, we present an interferometric scheme that extracts
\begin{equation}
    p\equiv\cos(\pi w)  
\end{equation}
and therefore distinguishes neighboring excited-state phases from each other: in the BA$'$ phase $p=1$, while in the TF$'$ and P$'$ phases $p=-1$. To measure $p$, first, an initial point $(n_0(0),\phi(0))$ on a trajectory at the $\xi$ and $\eta$ of interest is selected. Then the corresponding coherent state with $\phi_1=\phi_{-1}$, $|\psi(0)\rangle$, is prepared at $q=2|c|\xi$. The state freely evolves for the time $T$. Next, the spin states $m=0$ and $m=\pm 1$ are coupled by the internal-state beamsplitter $\exp(-i\frac{\pi}{2} \hat S_{\theta_0})$ with $\hat S_\theta \equiv \frac{1}{2}(\operatorname{e}^{-i\theta}\hat a_0^\dagger \hat g + \operatorname{e}^{i\theta} \hat g^\dagger \hat a_0)$, $\hat g\equiv (\hat a_1+\hat a_{-1})/\sqrt{2}$, and $\theta_0\equiv\pi/2-\phi(0)$. Finally, the expectation value of $\hat N_0/N$ is measured. In the mean-field limit, this yields~\cite{appendix}
\begin{equation}
    \label{eq:dop}
    \lim\limits_{N\to\infty}\frac{1}{N}\langle \psi(T)|\operatorname{e}^{i\frac{\pi}{2} \hat S_{{\theta_0}}}\hat N_0\operatorname{e}^{-i\frac{\pi}{2}\hat S_{{\theta_0}}}|\psi(T)\rangle=\frac{1-Vp}{2},
\end{equation}
where we have introduced the visibility $V=2\sqrt{1-n_0(0)}\sqrt{n_0(0)}$. As long as $n_0(0)\neq 0,1$, this unambiguously determines~$p$.


{\it Experimental realization.---}We detail the measurement of $p$ for $^{87}$Rb atoms in their hyperfine ground state~\cite{Luecke2011,Luo2017}. However, most of our discussion applies to any ferromagnetic spin-$1$ BEC. We assume that, initially, the condensate is in the state $\hat a_0^{\dagger N}|0\rangle/\sqrt{N!}$. Then a coherent state characterized by $n_0(0)$, $\phi(0)$, and $\phi_1=\phi_{-1}$ can be obtained by applying $\exp(-i\chi \hat S_{{\theta_0}})$ with $\cos^2(\chi/2)=n_0(0)$. Thus, both the state preparation and the beamsplitter are generated by $\hat S_{{\theta_0}}$ and can be implemented by a sequence of a phase shift $\exp(i({\theta_0}-\theta_{\rf})\hat N_0)$, a radio-frequency pulse $\exp(-i\zeta \hat S_{\theta_{\rf}})$~\footnote{The phase $\theta_{\rf}$ is fundamentally fixed by the interaction of the atoms with the radio-frequency pulse. While we do not state its value, it may be both derived and measured.} with $\zeta=\chi$ or $\zeta=\pi/2$, respectively, and another phase shift $\exp(-i({\theta_0}-\theta_{\rf})\hat N_0)$. Since we aim at the expectation value in Eq.~\eqref{eq:dop}, the first step of the state preparation and the last one of the beamsplitter can be omitted. $N_0$ can be measured, e.\,g., by a magnetic-field gradient that spatially separates the different spin states and subsequent absorptive imaging.
	
Reliably distinguishing $p=\pm 1$ requires a large visibility $V$, which can be maximized by choosing $n_0(0)$ as close to $1/2$ as possible. The optimal $n_0(0)$, $n_{\opt}$, is~\cite{appendix}:
\begin{equation}
    \label{eq:nopt}
    n_{\opt}=
    \begin{cases}
    \frac{1}{2}(1+\xi-\frac{\xi}{|\xi|}\sqrt{1+2\eta+\xi^2}) & \text{for }\eta<-\frac{1}{2}\\
    \frac{1}{2} &\text{for } -\frac{1}{2}\leq \eta\leq 0\\
    \frac{1}{2}(1-\frac{\eta}{\xi}) & \text{for } 0<\eta
    \end{cases}
\end{equation}
A corresponding $\phi(0)$, $\phi_{\opt}$, is obtained from
\begin{equation}
    \cos^2\!\phi_{\opt}=\frac{\xi(1-2n_{\opt})-\eta}{2n_{\opt}(1-n_{\opt})}.
\end{equation}
Figure~\ref{fig:2}a shows that the optimized visibility is large throughout the vast majority of the phase diagram. 
	
The coherence time in typical BEC experiments is limited to few seconds. This constrains the accessible periods $T$. 
It is known~\cite{Ueda2012, You2005, appendix} that 
\begin{equation}
    \label{eq:periodicity}
    \frac{|c|}{\hbar}T=
    \begin{cases}
    \sqrt{y}^{-1}K(x/y) & \text{for }\eta < \eta_*\\
    \sqrt{x}^{-1}K(y/x)& \text{for }\eta > \eta_*
    \end{cases},
\end{equation}
where $K(k^2)=\int_{0}^{\pi/2}\operatorname{d}\!\gamma\sqrt{1-k^2\sin^2\!\gamma}^{-1}$ is the complete elliptic integral of the first kind, $x=|\xi|\sqrt{1+\xi^2+2\eta}$, and $y=(x-\xi^2-\eta)/2$. $T$ diverges at the ESQPTs. Figure~\ref{fig:2}b displays $T$ for the typical interaction strength $|c|/\hbar=2\pi\times\SI{4}{Hz}$. Fortunately, $T$ exceeds a moderate value of, e.\,g., $\SI{0.3}{s}$ only in the immediate vicinity of the ESQPTs.
	
\begin{figure}[b]
\begin{center}
\includegraphics[scale=0.52]{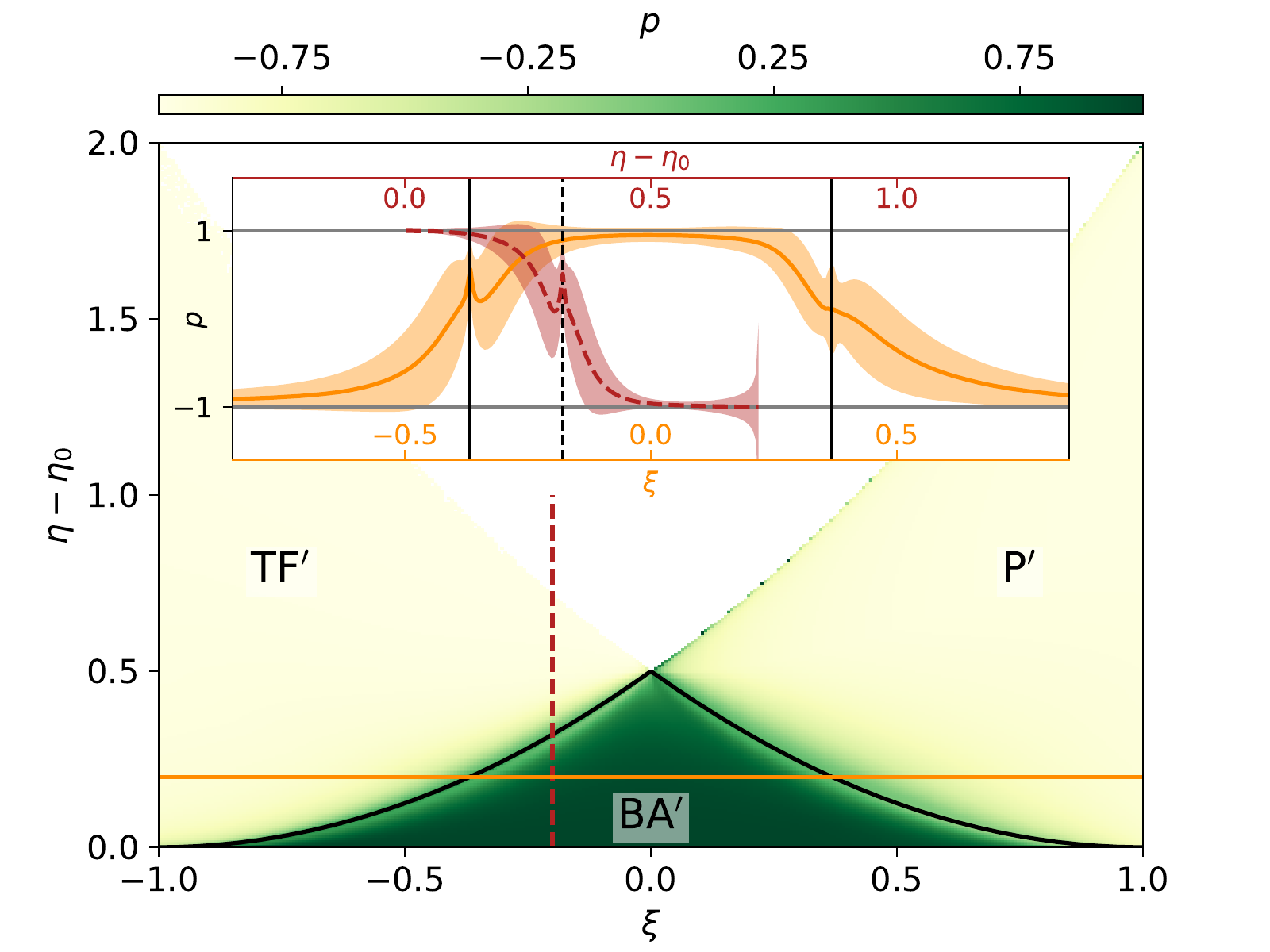}
\end{center}
\caption{Simulated measurement of $p$ for $N=100$ atoms. The finite-size results closely resemble the mean-field limit, where $p = 1$ in the BA$'$ phase and $p = -1$ in the TF$'$ and P$'$ phases. Black lines mark the ESQPTs. The inset shows $p$ along lines of constant $\xi=-0.2$ (red, dashed) and $\eta-\eta_0=0.2$ (orange, solid). The shaded regions indicate the standard deviation.}
\label{fig:3}
\end{figure}

So far we have considered only the mean-field limit, $N\to\infty$. 
To study the impact of a finite system size, we simulate a measurement of $p$ for $N=100$ bosons by exact diagonalization of the Hamiltonian density~\eqref{eq:HamiltonianQ}, see Fig.~\ref{fig:3}. 
The jump discontinuities signaling the ESQPTs in the mean-field limit are, as expected, smoothed at finite $N$. However, the BA$'$ phase can still be clearly distinguished from the TF$'$ and P$'$ phases. In typical experiments, $N$ is of the order of $10^4$ and, thus, a much better convergence to the mean-field limit can be expected.


{\it Conclusions.---}Ferromagnetic spin-$1$ BECs exhibit ESQPTs, which, in the mean-field limit, show up as a diverging DOS and a change in the topology of phase-space trajectories. We characterize the mean-field dynamics by a winding number $w$ that distinguishes the excited-state quantum phases from each other and, thus, is an order parameter. Adjacent phases differ in $|w|$ and can be told apart by interferometrically monitoring the coherent many-body dynamics in present-day experiments. Note that the local order parameter $\hat N_0/N$ that characterizes the ground-state QPTs~\cite{PhysRevLett.111.180401,Luo2017} cannot be directly generalized to excited states. The topological order parameter $w$, instead, is defined for all energies apart from the very ground state, where the trajectories reduce to single points.
Our results show that ESQPTs can be studied in well-controlled atomic quantum many-body systems, and that these studies are not limited to properties of the transition itself.
We propose a feasible experiment for characterizing excited-state quantum phases. This represents an important step towards employing ESQPTs in quantum state engineering.

Finally, we remark that our findings apply to any of the numerous quantum systems with the same mean-field limit, including bosonic two-level pairing models at zero generalized angular momentum~\cite{Iachello2008}. Our theoretical treatment of ESQPTs complements previous studies for opposite interaction sign~\cite{Iachello2008}. Bosonic two-level pairing models comprise, e.\,g., the LMG model, the vibron model for molecules, and the interacting boson model for nuclei.

\FloatBarrier
	
We thank Dmytro Bondarenko, Pavel Cejnar, Ignacio Cirac, and Reinhard Werner for valuable discussions. We acknowledge support by the Deutsche Forschungsgemeinschaft (DFG, German Research Foundation) under the SFB 1227 ``DQ-mat'', project A02, and under Germany's Excellence Strategy -- EXC-2123 QuantumFrontiers -- 390837967, and by the LabEx ENS-ICFP: ANR-10-LABX-0010/ANR-10-IDEX-0001-02 PSL*.
	

\onecolumngrid
\begin{appendix}

\section{Mean-field limit of bosonic systems}
\label{sec:MF}
We consider a system of $N\in\mathbb{N}$ (pseudo-)spin-$j$ bosons with $j\in \mathbb{N}_0/2$ or, equivalently, a system of $N$ bosons distributed among $2j+1\in\mathbb{N}$ modes. Such systems can be treated in terms of creation and annihilation operators $\annd_m$ and $\ann_m$, where $m\in\{j,j-1,\ldots,-j\}$ denotes the spin projection quantum number. Then $\hat N_m\equiv \annd_m \ann_m$ counts the number of particles in mode $m$. Our Hilbert space is restricted to eigenstates of $\hat N\equiv \sum_m\hat N_m$ with eigenvalue $N$.

We will focus on operators $\hat A$ with the following properties: 
\begin{enumerate}
	\item $\hat A$ is a polynomial in $\annd_m \ann_l/N$
	\item the polynomial coefficients $c_k(N)$ are time-independent 
	\item \label{property3} the $N$-dependence of the $c_k(N)$ is such that for any $k$ there are $d_k\in\mathbb{R}$ and $e_k\in\mathbb{C}$ with $|c_k(N)|\leq d_k\;\forall N$ and $\lim_{N\to\infty}c_k(N)=e_k$
\end{enumerate}
For example, the $c_k(N)$ may be independent of $N$ or include $\mathcal{O}(N^{-1})$ corrections. Note that $\hat A$ is simultaneously defined for all $N$ and $[\hat A,\hat N]=0$. We call the set of all such operators $\mathcal{A}$, and the subset of Hermitian operators $\mathcal{E}\subset\mathcal{A}$. Let $\hat H$ be the system's Hamiltonian and $\hat h\equiv \hat H/N$ the Hamiltonian density. We require that $\hat h\in\mathcal{E}$.

Let us discuss some further properties of $\mathcal{A}$. For any $\hat A,\hat B\in \mathcal{A}$:
\begin{enumerate}
	\setcounter{enumi}{3}
	\item Obviously, $\hat A\hat B\in\mathcal{A}$.
	\item $N[\hat A,\hat B]\in\mathcal{A}$. To confirm \textnumero~\ref{property3} of the defining properties, one may iteratively apply
	\begin{equation}
	    \label{eq:CommRules}
		[\hat D\hat E,\hat F]=[\hat D,\hat F]\hat E+\hat D[\hat E,\hat F]\;\;\text{and}\;\;[\hat D,\hat E\hat F]=[\hat D,\hat E]\hat F+\hat E[\hat D,\hat F],
	\end{equation}
	which holds for any operators $\hat D$, $\hat E$, and $\hat F$. This yields a finite number of terms, each of which contains a single elementary commutator, $[\ann_m/\sqrt{N},\annd_m/\sqrt{N}]=1/N$. For the coefficients of these terms, property \textnumero~\ref{property3} follows immediately.
	\item \label{norm} Let $\lVert\cdot\rVert_N$ denote the spectral norm in the $N$-particle Hilbert space. Then there is a $c\in\mathbb{R}$ such that $\lVert \hat A \rVert_N\leq c\;\forall N$. Note that $\lVert\cdot\rVert_N$ is sub-additive and sub-multiplicative, and that $\lVert\ann_m^{(\dagger)}\rVert_N=\sqrt{N}$. Hence, $c$ can be chosen to be the (finite) sum of the $d_k$ defined in property \textnumero~\ref{property3}.
\end{enumerate}

We employ the projective coherent states $|\bm{\alpha},N\rangle\equiv \frac{1}{\sqrt{N!}}(\sum_m\alpha_m \annd_m)^N|0\rangle$, where $\bm\alpha\in\mathbb{C}^{2j+1}$ comprises the $\alpha_m$, $\alpha_m\equiv\sqrt{n_m}\operatorname{e}^{i\phi_m}$, $n_m\geq 0$, $\phi_m\in[0,2\pi)$, and $\sum_m\!n_m=1$. These states are separable and fulfill $\ann_m|\bm{\alpha},N\rangle=\sqrt{N}\alpha_m|\bm{\alpha},N-1\rangle$ and $C_N\!\int\!\mathcal{D}\bm{\alpha}\,|\bm{\alpha},N\rangle\langle\bm{\alpha},N|=\mathbb{1}_N$ with $\mathcal{D}\bm{\alpha}\equiv\frac{1}{(2\pi)^3}\prod_m\!\operatorname{d}\!n_m\!\operatorname{d}\!\phi_m\,\delta\!\left(\sum_m\!n_m-1\!\right)$ and $C_N=\frac{(N+2j)!}{N!}$. $\mathbb{1}_N$ denotes the identity operator on the $N$-particle Hilbert space.

Let us now turn towards the mean-field limit $N\to\infty$. To start with, we consider the coherent-state expectation value of $\hat A\in\mathcal{A}$. Let $\!:\!\hat A\!:\!$ denote the normal ordering of $\hat A$. Using that $\ann_m|\bm{\alpha},N\rangle=\sqrt{N}\alpha_m|\bm{\alpha},N-1\rangle$ and, for any finite $k\in\mathbb{N}_0$, $\lim_{N\to\infty}\frac{N-k}{N}=1$, we obtain
$\lim_{N\to\infty} \langle \bm \alpha, N|\!:\!\hat A\!:\!|\bm\alpha, N\rangle$ from $\!:\!\hat A\!:\!$ by substituting the $\ann_m^{(\dagger)}$ by $\sqrt{N}\alpha_m^{(*)}$ and taking the limit $N\to\infty$. Note that, since the $\alpha_m$ commute, it does not matter whether we substitute the $\ann_m^{(\dagger)}$ in $:\!\hat A\!:$ or in $\hat A$. We denote the result by $A_{\mf}(\bm\alpha)$. From the scaling of $[\ann_m/\sqrt{N},\annd_m/\sqrt{N}]=1/N$ with $N$ we can conclude that $\lim_{N\to\infty} \langle \bm \alpha, N|\hat A -\!:\!\hat A\!:\!|\bm\alpha, N\rangle=0$. Hence, $ \lim_{N\to\infty} \langle \bm \alpha, N|\hat A|\bm\alpha, N\rangle= A_{\mf}(\bm \alpha)$. 

For $\hat A,\hat B\in \mathcal{A}$, obviously,
\begin{equation}
    \label{eq:abel}
    \lim_{N\to\infty} \langle \bm \alpha, N|\hat A\hat B|\bm\alpha, N\rangle=A_{\mf}(\bm \alpha)B_{\mf}(\bm \alpha).
\end{equation}
This is a central observation, which we can further generalize by means of \hyperref[tm:Tannery]{Tannery's} theorem, which we state below. Let us first show that
\begin{equation}
	\label{eq:T1}
	\lim_{N\to\infty} \langle \bm \alpha, N|\operatorname{e}^{z\hat A}|\bm\alpha, N\rangle =\operatorname{e}^{zA_{\mf}(\bm\alpha)}\;\;\forall z\in\mathbb{C}.
\end{equation}
We already know that  $\lim_{N\to\infty} \langle \bm \alpha, N|\hat A^k|\bm\alpha, N\rangle=A_{\mf}^k(\bm \alpha)\;\forall k\in\mathbb{N}_0$. \hyperref[tm:Tannery]{Tannery's} theorem ensures that we can pull the $N\to\infty$ limit into the exponential series. Its assumptions are fulfilled since, by property \textnumero~\ref{norm}, there is some $c\in\mathbb{R}$ such that $|\langle \bm \alpha, N|\hat A^k|\bm\alpha, N\rangle|\leq c^k\;\forall N$, and since $\sum_k \frac{|zc|^k}{k!}=\operatorname{e}^{|zc|}$ is finite. Note that this argument can be immediately generalized to arbitrary $N$-independent complex analytic functions on $\mathcal{A}$, the set of which we denote by $\mathcal{C}(\mathcal{A})$:
\begin{equation}
    \label{eq:Tanalytic}
    \lim_{N\to\infty} \langle \bm \alpha, N|f(\hat A)|\bm\alpha, N\rangle =f(A_{\mf}(\bm\alpha))\;\;\forall f\in\mathcal{C}(\mathcal{A}),\hat A\in\mathcal{A}.
\end{equation}

Similarly, employing the Baker-Campbell-Hausdorff formula, we obtain
\begin{equation}
	\label{eq:T2}
	\lim_{N\to\infty} \langle \bm \alpha, N|\operatorname{e}^{zN\hat A}\hat B \operatorname{e}^{-zN\hat A}|\bm\alpha, N\rangle
	= \sum_k \frac{z^k}{k!}K_{\mf}^{(k)}(\bm\alpha)\;\;\text{with}\;\;\hat K^{(0)}\equiv \hat B,\;\;\hat K^{(k)}\equiv N[\hat A,\hat K^{(k-1)}].
\end{equation}
The key step in proving Eq.~\eqref{eq:T2} is to demonstrate that we can find a suitable $N$-independent bound on $\lVert \hat K^{(k)}\rVert$. For $\lVert \hat A\lVert_N$ and $\lVert \hat B\lVert_N$, we first construct upper bounds $c_A$ and $c_B$ as suggested in the proof of property \textnumero~\ref{norm}. Then, iteratively applying the relation~\eqref{eq:CommRules}, we find that $\lVert \hat K^{(k)}\rVert \leq 2^{r_B} c_B(2^{r_A} c_A)^k$, where $r_A$ and $r_B$ are the polynomial degrees of $\hat A$ and $\hat B$, respectively. 

Finally, including a $\hat C\in\mathcal{A}$, we can show that
\begin{equation}
	\label{eq:T3}
	\lim_{N\to\infty} \langle \bm \alpha, N|\operatorname{e}^{zN\hat A}\hat B \hat C\operatorname{e}^{-zN\hat A}|\bm\alpha, N\rangle
	=\left(\lim_{N\to\infty} \langle \bm \alpha, N|\operatorname{e}^{zN\hat A}\hat B \operatorname{e}^{-zN\hat A}|\bm\alpha, N\rangle\right)\left( \lim_{N\to\infty}\langle \bm \alpha, N|\operatorname{e}^{zN\hat A} \hat C\operatorname{e}^{-zN\hat A}|\bm\alpha, N\rangle\right).
\end{equation}
Note that, in general, $\operatorname{e}^{zN\hat A}\hat B \operatorname{e}^{-zN\hat A}\notin\mathcal{A}$ and thus Eq.~\eqref{eq:T3} is not implied by Eq.~\eqref{eq:abel}. Instead, it can be derived from Eq.~\eqref{eq:T2} by using Eqs.~\eqref{eq:CommRules} and~\eqref{eq:abel} to observe that
\begin{equation}
	K_{\mf}^{(n)}[\hat B\hat C]=\sum_{l=0}^{n}\binom{n}{l}K_{\mf}^{(l)}[\hat B] K_{\mf}^{(n-l)}[\hat C],
\end{equation}
where the argument of $K_{\mf}^{(k)}$ specifies the operator $\hat K^{(0)}$, and $\hat K^{(k)}$ ensues inductively as defined in Eq.~\eqref{eq:T2}.

In the following two sections, we use Eqs.~\eqref{eq:T1},~\eqref{eq:T2}, and~\eqref{eq:T3} to derive the mean-field limits of the density of states and of the equations of motion. In Section~\ref{sec:theorems} we summarize, for completeness, some of the mathematical theorems we use.

\subsection{Density of states}
    \label{sec:DOS}
	We denote the energy per particle by $\epsilon$ and define the density of states (DOS) $\nu(\epsilon)$ by way of its Fourier transform:
	\begin{equation}
		\label{eq:DOSdef}
		\mathcal{F}[\nu](\zeta)\equiv\int\! \operatorname{d}\!\epsilon\,\operatorname{e}^{i \zeta \epsilon} \nu(\epsilon)\equiv\Tr \operatorname{e}^{i \zeta \hat h},\;\;\zeta\in\mathbb{R},
	\end{equation}
	where the trace is taken over the respective $N$-particle Hilbert space.
	To obtain the DOS in the mean-field limit, we argue that 
	\begin{align}
		\label{eq:DOSarg}
		\lim\limits_{N\to\infty}\frac{1}{N^{2j}}\Tr \operatorname{e}^{i \zeta \hat h}
		= \lim\limits_{N\to\infty} \frac{C_N}{N^{2j}} \int\! \mathcal{D}\bm{\alpha}\, \langle \bm \alpha, N|\operatorname{e}^{i\zeta \hat h}|\bm\alpha, N\rangle
		= \int\!\mathcal{D}\bm{\alpha}\, \operatorname{e}^{i\zeta h_{\mf}(\bm\alpha)}
		=\int\! \operatorname{d}\!\epsilon\,\operatorname{e}^{i \zeta \epsilon}\int\! \mathcal{D}\bm{\alpha}\,\delta(h_{\mf}(\bm\alpha)-\epsilon)
	\end{align}
	and conclude that
	\begin{equation}
		\label{eq:DOSconcl}
		\lim\limits_{N\to\infty}\frac{\nu(\epsilon)}{N^{2j}}=\int\!\mathcal{D}\bm{\alpha}\,\delta(h_{\mf}(\bm{\alpha})-\epsilon).	
	\end{equation}
	In the following we comment on some details of this derivation.
	
	First of all, note that $\nu(\epsilon)$ is well defined by Eq.~\eqref{eq:DOSdef}. For any $N$, the inverse Fourier transform of $\Tr\operatorname{e}^{i \zeta \hat h}$ is a unique tempered distribution, $\nu\in\mathcal{S}'(\mathbb{R})$.
	
	Next, we discuss each step of Eq.~\eqref{eq:DOSarg}. The first equality follows from the resolution of the identity in terms of coherent states: 
	\begin{equation}
		\Tr\hat D=\Tr\!\left[C_N\!\int\!\mathcal{D}\bm{\alpha}\,|\bm{\alpha},N\rangle\langle\bm{\alpha},N|\cdot\hat D\right]=C_N\! \int\! \mathcal{D}\bm{\alpha}\, \langle \bm \alpha, N|\hat D|\bm\alpha, N\rangle
	\end{equation}
	for any operator $\hat D$. The second equality in Eq.~\eqref{eq:DOSarg} comprises several steps. First, we note that $\lim_{N\to\infty} C_N/N^{2j}=1$. Second, we apply Eq.~\eqref{eq:T1} to the integrand:
	\begin{equation}
	    \label{eq:expConvergence}
		\lim\limits_{N\to\infty}\langle \bm \alpha, N|\operatorname{e}^{i\zeta \hat h}|\bm\alpha, N\rangle=\operatorname{e}^{i\zeta  h_{\mf}(\bm\alpha)}
	\end{equation}
	Third and last, we argue by \hyperref[tm:Lebesgue]{Lebesgue's} dominated convergence theorem that we can interchange the operation of taking the $N\to\infty$ limit with the integration. To check the assumptions of the theorem it is helpful to note that the domain of integration is compact, $\langle \bm \alpha, N|\operatorname{e}^{i\zeta \hat h}|\bm\alpha, N\rangle$ is a continuous function of $\bm\alpha$, and that $|\langle \bm \alpha, N|\operatorname{e}^{i\zeta \hat h}|\bm\alpha, N\rangle|\leq \lVert \operatorname{e}^{i\zeta \hat h}\rVert_N \leq 1\;\forall N$. The last step of Eq.~\eqref{eq:DOSarg} is, essentially, a change of variables. Some caution is needed at values of $\bm\alpha$ where the gradient of $h_{\mf}(\bm\alpha)$ vanishes. For measurable sets of $\bm\alpha$ with $h_{\mf}(\bm\alpha)=c$ the equality can be proven directly. Measure-zero sets with $\nabla h_{\mf}(\bm\alpha)=0$, e.\,g., isolated stationary points of $h_{\mf}$, can be excluded from the integration.
	
	Finally, to arrive at Eq.~\eqref{eq:DOSconcl} we demonstrate that, for any sequence of tempered distributions $f_N\in \mathcal{S}'(\mathbb{R})$, 
	\begin{equation}
		\label{eq:Fourier1}
		\lim_{N\to\infty}f_N=f\;\;\Leftrightarrow\;\;\lim_{N\to\infty}\mathcal{F}[f_N]=\mathcal{F}[f].
	\end{equation}
	Since the $f_N$ are distributions, we can demand convergence only in the following weak sense:
	\begin{equation}
		\label{eq:Fourier2}
		\lim_{N\to\infty}f_N=f\;\;\Leftrightarrow\;\; \lim_{N\to\infty}\int \!\operatorname{d}\!x\,f_N(x)t(x)= \int \!\operatorname{d}\!x\,f(x)t(x)
	\end{equation}
	for all test functions $t\in\mathcal{S}(\mathbb{R})$. Similarly, the Fourier transform of any $g\in \mathcal{S}'(\mathbb{R})$ is defined by
	\begin{equation}
		\label{eq:Fourier3}
	    \int \!\operatorname{d}\!x\,\mathcal{F}[g](x)t(x)\equiv \int \!\operatorname{d}\!x\,g(x)\mathcal{F}[t](x)\;\;\forall t\in\mathcal{S}(\mathbb{R}).
	\end{equation}
	Since the Fourier transformation is an automorphism on $\mathcal{S}(\mathbb{R})$, we can replace the arbitrary test function $t(x)$ in Eq.~\eqref{eq:Fourier2} by its Fourier transform $\mathcal{F}[t](x)$. This connects Eq.~\eqref{eq:Fourier2} with Eq.~\eqref{eq:Fourier3} and yields Eq.~\eqref{eq:Fourier1}.

\subsection{Equations of motion}
    \label{sec:EOM}
	We consider the Heisenberg representation $\hat A_{\rm{H}}(t)\equiv \operatorname{e}^{i\hat H t/\hbar}\hat A \operatorname{e}^{-i\hat H t/\hbar}$ of an operator $\hat A\in\mathcal{E}$. The Heisenberg equation of motion for $\langle \bm\alpha, N |\hat A_{\rm{H}}(t)|\bm\alpha,N\rangle$ reads
	\begin{equation}
		\label{eq:HB}
		\frac{\operatorname{d}}{\operatorname{d}\!t}\langle\bm\alpha,N|\hat A_{\rm{H}}(t)|\bm\alpha,N\rangle=\frac{i}{\hbar}\langle\bm\alpha,N|[\hat H,\hat A_{\rm{H}}(t)]|\bm\alpha,N\rangle.
	\end{equation}
	This section contains two results. We demonstrate that 
	\begin{equation}
	    \label{eq:DynLimit}
		\lim\limits_{N\to\infty}\langle\bm{\alpha},N|\hat A_{\rm{H}}(t)|\bm{\alpha},N\rangle=A_{\mf}(\bm{\alpha}_t),
	\end{equation}
	where $\bm{\alpha}_0\equiv \bm\alpha$ and $\bm\alpha_t$ consists of $\alpha_m(t)\equiv \sqrt{n_m(t)}\operatorname{e}^{i\phi_m(t)}$ with $n_m(t)\geq0$, $\phi_m(t)\in[0,2\pi)$, and $\sum_mn_m(t)=1$. Furthermore, we prove that the dynamics of $A_{\mf}(\bm{\alpha}_t)$ is governed by
	\begin{equation}
	    \label{eq:mfEOM}
	    \frac{\operatorname{d}}{\operatorname{d}\!t}A_{\mf}(\bm{\alpha}_t)=\frac{i}{\hbar}K_{\mf}(\bm{\alpha}_t)\;\;\text{with}\;\;\hat K\equiv [\hat H, \hat A].
	\end{equation}
	
	To derive Eq.~\eqref{eq:DynLimit}, we recall that $\hat A$ is a polynomial in $\annd_m \ann_l/N$ with coefficients $c_k(N)$. Let us define $\lambda_{ml}(t)\equiv\lim_{N\to\infty}\frac{1}{N}\langle\bm\alpha,N|[\annd_m \ann_l]_{\rm{H}}(t)|\bm\alpha,N\rangle$. Then, according to Eq.~\eqref{eq:T3}, $\lim_{N\to\infty}\langle\bm{\alpha},N|\hat A_{\rm{H}}(t)|\bm{\alpha},N\rangle$ is a polynomial in $\lambda_{ml}(t)$ with the respective coefficients $\lim_{N\to\infty}c_k(N)$. Next, we argue that $\lambda_{ml}(t)$ can be parametrized, without loss of generality, by $\sqrt{n_m(t)n_l(t)}\operatorname{e}^{-i(\phi_m(t)-\phi_l(t))}$ with $n_m(t)$ and $\phi_m(t)$ as introduced above. At $t=0$ this parametrization is obviously correct. For $m=l$, $\lambda_{mm}(t)=n_m(t)$ is a valid parametrization because $\lambda_{mm}(t)\geq 0$ and $\sum_m\lambda_{mm}(t)=\lim_{N\to\infty}\frac{1}{N}\langle\bm\alpha,N|\hat N_{\rm{H}}(t)|\bm\alpha,N\rangle=1$.
	Employing, again, Eq.~\eqref{eq:T3}, we find
	\begin{align} 
	    \begin{split}
    	\lambda_{ml}\lambda_{lm}&=\lim\limits_{N\to\infty}\frac{1}{N^2}\langle\bm\alpha,N|[\annd_m\ann_l\annd_l\ann_m]_{\rm{H}}(t)|\bm\alpha,N\rangle\\
    	&=\lim\limits_{N\to\infty}\frac{1}{N^2}\langle\bm\alpha,N|[\annd_m \ann_m\annd_l\ann_l+ \annd_m\ann_m]_{\rm{H}}(t)|\bm\alpha,N\rangle=n_m(t)n_l(t).
    	\end{split}
	\end{align} 
	Together with $\lambda_{ml}(t)=\lambda_{lm}^*(t)$ this entails $|\lambda_{ml}(t)|=|\lambda_{lm}(t)|=\sqrt{n_m(t)n_l(t)}$. The phases of all $\lambda_{ml}(t)$ with $m\neq l$ can be deduced from the phases of $\lambda_{1l}(t)$ by using the relations $\lambda_{ml}(t)=\lambda_{lm}^*(t)$ and $\lambda_{ml}(t)\lambda_{lk}(t)\lambda_{km}(t)=n_m(t)n_l(t)n_k(t)\in\mathbb{R}$. The parametrization $\lambda_{ml}(t)=\sqrt{n_m(t)n_l(t)}\operatorname{e}^{-i(\phi_m(t)-\phi_l(t))}$ reflects these relations without constraining the $\lambda_{ml}$ any further.
	
	To obtain the mean-field equation of motion~\eqref{eq:mfEOM}, we take the $N\to\infty$ limit of Eq.~\eqref{eq:HB}. In any time interval $[t_1,t_2]$, Theorem~\ref{tm:limitderivative} permits to interchange the limit with the time derivative because the right-hand side (RHS) of Eq.~\eqref{eq:HB} is continuous in $t$ and uniformly converges for $N\to \infty$. Let us prove the uniform convergence. We know from Eq.~\eqref{eq:T2} that the pointwise limit of
	\begin{equation}
		f_N(t)\equiv\langle\bm\alpha,N|[\hat H, \hat A_{\rm{H}}(t)]|\bm\alpha,N\rangle=\sum_k \frac{(it/\hbar)^k}{k!}\langle\bm\alpha,N|\hat K^{(k+1)}|\bm\alpha,N\rangle\;\;\text{with}\;\;\hat K^{0}\equiv \hat A,\;\;\hat K^{k}\equiv [\hat H,\hat K^{(k-1)}]
	\end{equation}
	is
	\begin{equation}
		\lim_{N\to\infty}f_N(t)\equiv f(t)=\sum_k \frac{(it/\hbar)^k}{k!}K^{(k+1)}_{\mf}.
	\end{equation}
	It is sufficient to show that the RHS of
	\begin{equation}
		\label{eq:uniformConv}
		|f_N(t)-f(t)|\leq \sum_k \frac{|t/\hbar|^k}{k!}\left|\langle\bm\alpha,N|\hat K^{(k+1)}|\bm\alpha,N\rangle-K_{\mf}^{(k+1)}(\bm\alpha)\right|,\;\;t\in[t_1,t_2]
	\end{equation} 
	uniformly converges to zero as $N\to\infty$. Similarly to the proof of Eq.~\eqref{eq:T2}, we can find some $c,\tilde{c}\in\mathbb{R}$ such that
	\begin{equation}
		|\langle\bm\alpha,N|\hat K^{(k+1)}|\bm\alpha,N\rangle|\leq \tilde{c}c^k\;\forall N\;\; \Rightarrow\;\;\left|\langle\bm\alpha,N|\hat K^{(k+1)}|\bm\alpha,N\rangle-K_{\mf}^{(k+1)}(\bm\alpha)\right|\leq 2\tilde{c}c^k\;\forall N.
	\end{equation}  
	Hence, we can apply \hyperref[tm:Tannery]{Tannery's} theorem, which yields that the RHS of Eq.~\eqref{eq:uniformConv} converges to zero pointwise. The RHS of Eq.~\eqref{eq:uniformConv} is a strictly increasing function of $|t|$. Let us assume, without loss of generality, that $|t_2|\geq |t_1|$. Then, for any $N$, the RHS of Eq.~\eqref{eq:uniformConv} is absolutely bounded by its value at $t_2$ and the pointwise convergence to zero in $t_2$ implies uniform convergence. 
		
\subsection{Mathematical supplement}
	\label{sec:theorems}
 	For completeness, we state here some well-known theorems which we have used above:
	\begin{theorem}[Tannery \cite{Loya}]
	    \label{tm:Tannery}
	    Consider the sequence $a_k(n)\in\mathbb{C}$ with $k\in\mathbb{N}_0, n\in\mathbb{N}$ and assume that for any $k$ there are $b_k,c_k$ such that $\lim_{n\to\infty} a_k(n)=b_k$, $|a_k(n)|\leq c_k$ $\forall n$, and $\sum_k c_k<\infty$. Then $\lim_{n\to\infty}\sum_k a_k(n)=\sum_k b_k$. 
	\end{theorem}
	\begin{theorem}[Lebesgue \cite{Koenigsberger}]
	    \label{tm:Lebesgue}
	    Let $f_n:U\subset\mathbb{R}^d\to\mathbb{C}$, $n\in\mathbb{N}$ be Lebesgue integrable functions which, for $n\to\infty$, converge pointwise to a function f and are dominated by some Lebesgue integrable function g, i.\,e., $|f_n(x)|\leq g(x)$ $\forall n\in\mathbb{N}, x\in U$.
	    Then $f$ is integrable and
	    \begin{equation}
	        \lim\limits_{n\to\infty}\int_U\!\operatorname{d}\!x \,f_n(x) = \int_U\!\operatorname{d}\!x \,f(x).
	    \end{equation}
	\end{theorem}
	\begin{theorem}[\cite{Forster}]
	    \label{tm:limitderivative}
	    Let $f_n:[a,b]\to\mathbb{R}$, $n\in\mathbb{N}$ be continuously differentiable functions which, for $n\to\infty$, converge pointwise to $f$. Let the sequence of derivatives $f'_n:[a,b]\to\mathbb{R}$ converge uniformly. Then $f$ is differentiable and 
	    \begin{equation}
	        f'(x)=\lim_{n\to\infty}f'_n(x)\;\;\forall x\in[a,b].
	    \end{equation}
	\end{theorem}
	
\section{Ferromagnetic spin-1 Bose-Einstein condensate with zero magnetization}
    \label{sec:spin1}
	In this section we focus on a ferromagnetic spin-1 Bose-Einstein condensate (BEC), which we model by the Hamiltonian density in Eq.~\eqref{eq:HamiltonianQ} from the main text,
	\begin{align}
	    \begin{split}
		\label{eq:HamiltonianSpin1}
		\hat h = \frac{q}{N}\left(\frac{1}{2}N-\hat N_0\right)+\frac{c}{N^2}\left[\vphantom{\frac{1}{2}}\annd_1 \annd_{-1} \ann_0^2+\ann_0^{\dagger 2}\ann_1 \ann_{-1}\right.+\left.\hat N_0\left(N-\hat N_0+\frac{1}{2}\right) + \frac{1}{2}\hat D^2\right].
	    \end{split}
	\end{align}
	Recall that $q\in\mathbb{R}$, $c<0$, $\hat D=\hat N_1-\hat N_{-1}$ is the magnetization, and $[\hat h, \hat D]=0$. We are particularly interested in the case of zero magnetization. This has several reasons. First, the present work is motivated by the utility of ground-state quantum phase transitions (QPTs) in the magnetization-free subspace~\cite{Pezze2019,Feldmann2018}. Second, previous results on excited-state QPTs (ESQPTs) suggest~\cite{Iachello2008} that the signatures should be most pronounced at zero magnetization. Third, the restriction to zero magnetization eases the computations.
	
	As discussed in Section~\ref{sec:MF}, we base our mean-field study on spin-1 projective coherent states. Most of these states are no eigenstates of $\hat D$. It is therefore not obvious how to restrict the $N$-particle Hilbert space to the eigenspace of $\hat D$ with eigenvalue $D=0$. In Section~\ref{sec:restrDOS} we demonstrate that the mean-field limit of the restricted density of states can be still expressed in terms of spin-1 coherent states. 
	
	For the mean-field dynamics of expectation values, we simply confine ourselves to coherent states $|\bm\alpha,N\rangle$ with $\langle\bm\alpha,N|\hat D |\bm\alpha,N\rangle=0$ or, equivalently, with $|\alpha_1|^2=|\alpha_{-1}|^2$. Importantly, such states can be readily realized experimentally. To identify the mean-field Hamiltonian $h_{\mf}(\bm\alpha)=\lim_{N\to\infty}\langle\bm{\alpha},N|\hat h|\bm{\alpha},N\rangle$, we substitute the $\ann_m^{(\dagger)}$ by $\sqrt{N}\alpha_m^{(*)}$ with $\alpha_m\equiv\sqrt{n_m}\operatorname{e}^{i\phi_m}$ and take the limit $N\to\infty$, as explained in the introduction to Section~\ref{sec:MF}. We introduce $\phi\equiv \phi_0- (\phi_1 + \phi_{-1})/2$ and $\xi\equiv \frac{q}{2|c|}$ and, for $n_1=n_{-1}$, obtain
	\begin{equation}
		\label{eq:ClassicalModel1}
		\frac{h_{\mf}}{|c|} =\xi(1-2n_0)-2n_0(1-n_0)\cos^2\!\phi,
	\end{equation}
	cf. Eq.~\eqref{eq:HamiltonianC} from the main text. Applying Eq.~\eqref{eq:mfEOM} to $\hat N_0/N$, $(\annd_1 \annd_{-1} \ann_0^2+\ann_0^{\dagger 2}\ann_1\ann_{-1})/N^2$, $\annd_1 \ann_{-1}/N$, and $\annd_{-1} \ann_1/N$ yields the equations of motion in Eq.~\eqref{eq:EOMexplicit} from the main text:
	\begin{equation}
		\label{eq:ClassicalModel2}
	    \frac{\operatorname{d}}{\operatorname{d}\!\tau}n_0=\frac{\partial}{\partial\phi}\frac{h_{\mf}}{|c|},\;\; \frac{\operatorname{d}}{\operatorname{d}\!\tau}\phi=-\frac{\partial}{\partial n_0}\frac{h_{\mf}}{|c|}, \;\;\text{and}\;\;\frac{\operatorname{d}}{\operatorname{d}\!\tau}(\phi_1-\phi_{-1})=0
	\end{equation}
	The first two equations of motion are Hamilton's equations for the Hamiltonian $h_{\mf}$ and the canonical coordinates $n_0$ and $\phi$.
	The corresponding phase space is a sphere with $z$-axis $0\leq n_0\leq 1$ and azimuthal angle $\phi\in[0,2\pi)$. Below, Section~\ref{sec:orbits} provides the classical phase-space trajectories. In Section~\ref{sec:dynamics} we review the dynamics of $n_0$.
	
	In the main text, we have introduced an order parameter for the ESQPTs in ferromagnetic spin-1 BECs with zero magnetization and have proposed to reveal this order parameter by interferometry.
	In Section~\ref{sec:visibility} we supplement some mathematical details regarding our measurement prescription.
	
\subsection{Restricting the density of states}
    \label{sec:restrDOS}
     The Fock basis of the $N$-particle Hilbert space consists of the joint eigenstates $|N_1,N_0,N_{-1}\rangle$ of the $\hat N_m$ with eigenvalues $N_m$ and $\sum_m N_m=N$. In this basis, the projection onto the eigenspace of $\hat D$ with eigenvalue $D=0$ reads
    \begin{equation}
        \hat{\mathbb{P}}_0=\sum_{k=0}^{N/2}|k,N-2k,k\rangle\langle k,N-2k,k|.
    \end{equation}
    We define the density of states (DOS) in the $D=0$ subspace by
	\begin{equation}
		\label{eq:rDOSdef}
		\mathcal{F}[\nu_0](\zeta)=\int\! \operatorname{d}\!\epsilon\,\operatorname{e}^{i \zeta \epsilon} \nu_0(\epsilon)\equiv\Tr \hat{\mathbb{P}}_0\operatorname{e}^{i \zeta \hat h},\;\;\zeta\in\mathbb{R}.
	\end{equation}  
	Below, we show that
	\begin{equation}
	    \label{eq:rDOSarg}
	    \lim\limits_{N\to\infty}\frac{1}{N}\Tr \hat{\mathbb{P}}_0\operatorname{e}^{i \zeta \hat h}
		= \int\! \mathcal{D}\bm{\alpha}\delta(n_1-n_{-1})\, \operatorname{e}^{i\zeta h_{\mf}(\bm\alpha)}.
	\end{equation}
	In the same way as in Section~\ref{sec:DOS}, this implies 
	\begin{equation}
	    \lim\limits_{N\to\infty}\frac{\nu_0(\epsilon)}{N}=\int\!\mathcal{D}\bm{\alpha}\delta(n_1-n_{-1})\,\delta(h_{\mf}(\bm{\alpha})-\epsilon)
	\end{equation}
	for the restricted DOS in the mean-field limit.
	Expressing $\nu_0$ as a function of $\eta=\epsilon/|c|$ yields Eq.~\eqref{eq:DOSReduced} from the main text.
	
    In the following, we prove that
	\begin{equation}
	    \label{eq:rDOSgen}
	    \lim\limits_{N\to\infty}\frac{1}{N}\Tr \hat{\mathbb{P}}_0 f(\hat A)
		= \int\! \mathcal{D}\bm{\alpha}\delta(n_1-n_{-1})\, f(A_{\mf}(\bm\alpha))\;\;\forall f\in\mathcal{C}(\mathcal{A}),\hat A\in\mathcal{A},
	\end{equation}
	where, as before, $\mathcal{D}\bm{\alpha}\equiv\frac{1}{(2\pi)^3}\prod_m\!\operatorname{d}\!n_m\!\operatorname{d}\!\phi_m\,\delta\!\left(\sum_m\!n_m-1\!\right)$ and $\mathcal{C}(\mathcal{A})$ denotes the complex analytic functions on $\mathcal{A}$. Equation~\eqref{eq:rDOSarg} immediately follows as a special case. 
	First, we assume that
	\begin{equation}
	    \label{eq:rDOSass}
	    \lim\limits_{N\to\infty}\frac{1}{N}\Tr \hat{\mathbb{P}}_0 f(\hat A) = \int\! \prod_m\operatorname{d}\!n_m\!\operatorname{d}\!\phi_m\mu(\bm\alpha)\,f(A_{\mf}(\bm\alpha))
	\end{equation}
    and determine $\mu(\bm\alpha)$. We observe that $\hat{\mathbb{P}}_0$ is invariant under phase shifts:
    \begin{equation}
        \operatorname{e}^{-i\theta \hat N_m}\hat{\mathbb{P}}_0\operatorname{e}^{i\theta \hat N_m}=\hat{\mathbb{P}}_0\;\Rightarrow\;\Tr \hat{\mathbb{P}}_0 f(\operatorname{e}^{i\theta \hat N_m}\hat A\operatorname{e}^{-i\theta \hat N_m})=\Tr \hat{\mathbb{P}}_0 f(\hat A)\;\;\forall \theta, m
    \end{equation}
    The mean-field limit of $\operatorname{e}^{i\theta \hat N_m}\hat A\operatorname{e}^{-i\theta \hat N_m}$ is $A_{\mf}(\bm\alpha')$ with $\phi_m'=\phi_m-\theta$ and, apart from that, $\bm\alpha'$ coinciding with $\bm\alpha$. Hence, $\mu(\bm\alpha)$ cannot depend on any of the $\phi_m$. For an arbitrary function $f\in\mathcal{C}(\mathcal{A})$ of $(\hat N_1 + \hat N_0+\hat N_{-1})/N$,
	\begin{align}
	    \lim\limits_{N\to\infty}\frac{1}{N}\Tr \hat{\mathbb{P}}_0 f\!\left(\frac{\hat N_1 + \hat N_0+\hat{N}_{-1}}{N}\right)=\lim\limits_{N\to\infty}\frac{1}{N}\sum_{k=0}^{N/2} f(1)=\frac{1}{2}f(1).
	\end{align}
	Similarly, for functions of $\hat D/N$,
	\begin{align}
	    \lim\limits_{N\to\infty}\frac{1}{N}\Tr \hat{\mathbb{P}}_0 f\!\left(\frac{\hat D}{N}\right)=\frac{1}{2}f(0).
	\end{align}
	Hence, $\mu(\bm\alpha)=\tilde\mu(n_0)\delta(\sum_m n_m -1)\delta(n_1-n_{-1})$. To determine $\tilde\mu(n_0)$, we consider functions of $\hat N_0/N$:
	\begin{align}
	    \lim\limits_{N\to\infty}\frac{1}{N}\Tr \hat{\mathbb{P}}_0 f\!\left(\frac{\hat N_0}{N}\right)= \lim\limits_{N\to\infty}\frac{1}{N}\sum_{k=0}^{N/2}f\!\left(\frac{N-2k}{N}\right)=\frac{1}{2}\int\!\operatorname{d}\!n_0\,f(n_0).
	\end{align}
	This, finally, yields $\mu(\bm\alpha)=\frac{1}{(2\pi)^3}\delta(\sum_m n_m -1)\delta(n_1-n_{-1})$ as required for Eq.~\eqref{eq:rDOSgen}.
	
	The proof of our assumption, Eq.~\eqref{eq:rDOSass}, relies on results from Ref.~\cite{WernerRaggio89}. Recall that all $\hat A\in\mathcal{A}$ and $f(\hat A)$ with $f\in\mathcal{C}(\mathcal{A})$ are endomorphisms on $N$-particle Hilbert spaces with arbitrary $N$. Considered as sequences in $N$, they belong to the set of approximately symmetric sequences $\tilde{\mathcal{Y}}$ defined in Ref.~\cite{WernerRaggio89}. On each $N$-particle Hilbert space, we introduce the state $\hat \rho_N\equiv \frac{1}{\lfloor N/2\rfloor+1}\hat{\mathbb{P}}_0$ and the corresponding linear functional $T_N:\tilde{\mathcal{Y}}\to \mathbb{C}$, $T_N(\hat Y)\equiv\Tr\hat\rho_N\hat Y_N$, where $\hat Y_N$ denotes the sequence elements of $\hat Y$. The $T_N$ constitute a sequence $T_\mathbb{N}$ in a compact space~\cite{WernerRaggio89}. The compactness has two consequences. First, $T_\mathbb{N}$ has a convergent subsequence. Second, if all convergent subsequences of $T_\mathbb{N}$ converge to the same $T_\infty$, so does the entire $T_\mathbb{N}$. According to Propositions III.3 and IV.5 in Ref.~\cite{WernerRaggio89}, the limit of any convergent subsequence of $T_\mathbb{N}$ assumes the form
	\begin{equation}
	    T_\infty(\hat Y)=\int\! \prod_m\operatorname{d}\!n_m\!\operatorname{d}\!\phi_m\mu(\bm\alpha)\lim\limits_{N\to\infty} \langle \bm \alpha, N|\hat Y|\bm\alpha, N\rangle.
	\end{equation}
	The arguments from the previous paragraph immediately yield $\mu(\bm\alpha)=\frac{2}{(2\pi)^3}\delta(\sum_m n_m -1)\delta(n_1-n_{-1})$, where the additional factor of 2 reflects the different normalization of $\hat \rho_N$ and $\frac{1}{N}\hat{\mathbb{P}}_0$. Hence, $T_\infty$ does not depend on the convergent subsequence under consideration and $T_\mathbb{N}$ converges to $T_\infty$. Finally, we observe that 
	\begin{equation}
	    \lim\limits_{N\to\infty}\frac{1}{N}\Tr \hat{\mathbb{P}}_0 f(\hat A)=\frac{1}{2}T_\infty(f(\hat A)).
	\end{equation}
	Recalling that $ \lim\limits_{N\to\infty} \langle \bm \alpha, N|f(\hat A)|\bm\alpha, N\rangle=f(A_{\mf}(\bm\alpha))$, see Eq.~\eqref{eq:Tanalytic}, completes the proof.
	
\subsection{Phase-space trajectories}
\label{sec:orbits}
	An energy hypersurface at $\xi$ and $\eta$ consists of all phase-space points $(n_0,\phi)$ which fulfill
	\begin{equation}
		\label{eq:implicitEHS}
		\eta=\frac{h_{\mf}}{|c|}=\xi(1-2n_0)-2n_0(1-n_0)\cos^2\!\phi.
	\end{equation}
	For $0<|\xi|<1$, Eq.~\eqref{eq:implicitEHS} can be rewritten as
	\begin{equation}
	    \label{eq:explicitEHS}
	    n_0(\phi) = 
	    \begin{cases}
	    m_\pm(\phi) \;\;\forall \cos^2\!\phi\geq\sqrt{\eta^2-\xi^2}-\eta
	    &\text{for}\;\;\eta < \eta_*\\
        \begin{cases}
        m_+(\phi)\;\;\forall \phi&\text{for}\;\;\xi<0 \\
        m_-(\phi)\;\;\forall \phi&\text{for}\;\;\xi>0
        \end{cases} &\text{for}\;\; \eta>\eta_*\\
        \frac{1}{2}\!\left(\!1-\frac{\xi}{|\xi|}\!\right)+\frac{\xi}{\cos^2\!\phi} \;\;\forall \cos^2\!\phi\geq|\xi| &\text{for}\;\;\eta=\eta_*
	    \end{cases}
	\end{equation}
	with $m_\pm(\phi)=\frac{1}{2\cos^2\!\phi}(\cos^2\!\phi+\xi\pm \sqrt{\Delta})$ and $\Delta=\cos^4\!\phi+2\eta\cos^2\!\phi+\xi^2$. Recall that $\eta_0=-\frac{1}{2}(\xi^2+1)$, $\eta_0\leq\eta\leq|\xi|$, and $\eta_*=-|\xi|$. 
	
	A phase-space trajectory is the set of all points $(n_0,\phi)$ which are connected by the Hamiltonian dynamics. Particularly, any closed line of constant $\eta$ which does not pass through a stationary point of $h_{\mf}$ is a trajectory.
	For each $\eta>\eta_*$, the energy hypersurface is a closed line by itself, while for $\eta_0<\eta<\eta_*$ each energy hypersurface comprises two disconnected closed lines. Since the stationary points of $h_{\mf}$ are at $\eta_0$ and $\eta_*$, we conclude that the phase-space trajectories for $\eta\notin \{\eta_0,\eta_*\}$ are the connected components of the energy hypersurfaces~\eqref{eq:explicitEHS}. This result can be extended to $\eta_0$, where the energy hypersurface consists of two stationary points, each of which is its own trajectory. At $\eta_*$, the energy hypersurface has the shape of an eight with the stationary point located at the intersection. There are, hence, three trajectories: the two wings of the eight excluding the stationary point, and the stationary point itself.
	
	Phase-space trajectories are commonly assigned the direction in which they are traced by the evolution forward in time. This direction is determined by the equations of motion, see Eq.~\eqref{eq:ClassicalModel2}.
	
\subsection{Dynamics}
    \label{sec:dynamics}
	The dynamics of $n_0$ is governed by 
	\begin{equation}
		\label{eq:EOMn0}
		\frac{\operatorname{d}}{\operatorname{d}\!\tau}n_0=\frac{\partial}{\partial \phi}\frac{h_{\mf}}{|c|}=4n_0(1-n_0)\cos\phi\sin\phi.
	\end{equation}
	We square Eq.~\eqref{eq:EOMn0} and, exploiting the conservation of $\eta$, cf. Eq.~\eqref{eq:implicitEHS}, obtain for $\xi\neq0$
	\begin{equation}
		\label{eq:EOMn0Sq}
		\left(\frac{\operatorname{d}}{\operatorname{d}\!\tau}n_0\right)^2=16\xi(n_0-z_0)(n_0-z_+)(n_0-z_-)
	\end{equation}
	with $z_0=\frac{1}{2}(1-\eta/\xi)$ and $z_\pm=\frac{1}{2}(1+\xi\pm\sqrt{1+\xi^2+2\eta})$. Recall that ESQPTs at $\eta_*=-|\xi|$ and $0<|\xi|<1$ divide the $\xi$-$\eta$-plane into three excited-state quantum phases: the TF$'$ phase for $\eta > \eta_*$ and $\xi<0$, the P$'$ phase for $\eta>\eta_*$ and $\xi>0$, and the BA$'$ phase for $\eta<\eta_*$. In the TF$'$ and P$'$ phases $z_-\leq z_0\leq z_+$, while in the BA$'$ phase $z_0\leq z_-\leq z_+$ for $\xi<0$ and $z_-\leq z_+\leq z_0$ for $\xi>0$. 
	
	Let us introduce $x_1=\frac{1}{2}(1+|\xi|-\sqrt{1+\xi^2+2\eta})$ and
	\begin{equation}
	    x_2=
	    \begin{cases}
	    \frac{1}{2}(1+|\xi|+\sqrt{1+\xi^2+2\eta}) & \text{for }\eta\leq\eta_*\\
	    \frac{1}{2}(1-\eta/|\xi|) & \text{for }\eta>\eta_*
	    \end{cases},\quad
	    x_3=
	    \begin{cases}
	    \frac{1}{2}(1-\eta/|\xi|) & \text{for }\eta\leq\eta_*\\
	    \frac{1}{2}(1+|\xi|+\sqrt{1+\xi^2+2\eta}) & \text{for }\eta>\eta_*
	    \end{cases}.
	\end{equation}
	Note that $x_1\leq x_2\leq x_3$ and that for $\xi>0$ the $x_i$ coincide with the appropriately ordered zeroes $z_0$ and $z_\pm$. According to Refs.~\cite{You2005,Ueda2012},
	\begin{equation}
		\label{eq:DYNn0}
		n_0(\tau)=x_2-(x_2-x_1)\operatorname{cn}^2\!\left(2\sqrt{|\xi|(x_3-x_1)}\tau+u_0,\frac{x_2-x_1}{x_3-x_1}\right)\equiv \tilde n_0(\tau)\;\;\text{for}\;\;\xi>0,
	\end{equation}
	where $\operatorname{cn}(u;k^2)$ is the Jacobi elliptic cosine and $u_0$ accounts for the initial conditions\footnote{\mbox{Note that at $(|\xi|=1,\eta=-1)$ the denominator $x_3-x_1$ vanishes. Hence, $\tilde n_0(\tau)$ has to be computed by taking an appropriate limit.}}. It can be easily verified that Eq.~\eqref{eq:DYNn0} solves Eq.~\eqref{eq:EOMn0Sq}. The dynamics for $\xi<0$ and $\xi>0$ are related by $h_{\mf}(\xi,n_0,\phi)=h_{\mf}(-\xi,1-n_0,\phi)$. Combining this with the equations of motion in Eq.~\eqref{eq:ClassicalModel2} and using the time-reversal symmetry of $\tilde n_0(\tau)$ one can show that
	\begin{equation}
		n_0(\tau)=1-\tilde n_0(\tau)\;\;\text{for}\;\;\xi<0.
	\end{equation}
	
	The evolution $n_0(\tau)$ is periodic with period
	\begin{equation}
		\mathcal{T} =\frac{1}{\sqrt{|\xi|(x_3-x_1)}}K\!\left(\frac{x_2-x_1}{x_3-x_1}\right),
	\end{equation}
	where $K(k^2)=\int_{0}^{\pi/2}\operatorname{d}\!\gamma\sqrt{1-k^2\sin^2\!\gamma}^{-1}$ is the complete elliptic integral of the first kind. Plugging in the respective expressions for the $x_i$ yields 
	\begin{equation}
		\mathcal T=
		\begin{cases}
		\sqrt{y}^{-1}K(x/y) & \text{for }\eta < \eta_*\\
		\sqrt{x}^{-1}K(y/x)& \text{for }\eta > \eta_*
		\end{cases}
	\end{equation} 
	with $x=|\xi|\sqrt{1+\xi^2+2\eta}$ and $y=(x-\xi^2-\eta)/2$. To obtain Eq.~\eqref{eq:periodicity} from the main text, recall that $\tau = |c|t/\hbar$. The periodicity diverges at the ESQPTs, as can be derived from $\lim_{k^2\nearrow 1}K(k^2)=\infty$.
	
\subsection{\texorpdfstring{Measuring the order parameter $w$}{Measuring the order parameter w}}
    \label{sec:visibility}
    In the main text, we have introduced the order parameter $w$, which distinguishes between the TF$'$ ($w=-1$), the P$'$ ($w=1$), and the BA$'$ ($w=0$) phase. Our measurement prescription for $p=\cos(\pi w)$ is summarized in Eq.~\eqref{eq:dop} from the main text. 
    It relies on the mean-field dynamics at a given $\xi$ and $\eta$. To evaluate the corresponding mean-field limit, we have employed Eq.~\eqref{eq:DynLimit}.

    The visibility $V=2\sqrt{1-n_0(0)}\sqrt{n_0(0)}$, which depends on the initial condition $n_0(0)$, quantifies how well one can experimentally tell $|w|=1$ from $|w|=0$ and, thus, neighboring excited-state quantum phases from each other. To optimize $V$, $n_0(0)$ has to be chosen as close to $1/2$ as is compatible with the periodic dynamics $n_0(\tau)$ at the $\xi$ and $\eta$ under consideration. We denote the optimal value of $n_0(0)$ by $n_{\rm{opt}}$.
    
    According to Section~\ref{sec:dynamics}, $n_0(\tau)$ oscillates in the TF$'$ phase between the minimum value $z_0$ and the maximum value $z_+$, in the P$'$ phase between the minimum $z_-$ and the maximum $z_0$, and in the BA$'$ phase between $z_-$ and $z_+$. We observe that $z_0\leq 1/2\Leftrightarrow \eta/\xi\geq 0$ and $z_0\geq 1/2\Leftrightarrow \eta/\xi\leq 0$. Furthermore,  for $\xi<0$ it is obvious that $z_-<1/2$, and for $\xi>0$ that $z_+>1/2$. Finally, one can show for $\xi>0$ that $z_-\leq 1/2\Leftrightarrow \eta\geq-1/2$ and for $\xi<0$ that $z_+\geq 1/2\Leftrightarrow \eta\geq -1/2$. Combining these findings yields Eq.~\eqref{eq:nopt} from the main text:
    \begin{equation}
        n_{\rm{opt}}=
        \begin{cases}
        \frac{1}{2}(1+\xi-\frac{\xi}{|\xi|}\sqrt{1+2\eta+\xi^2}) & \text{for }\eta<-1/2\\
        \frac{1}{2} &\text{for } -1/2\leq \eta\leq 0\\
        \frac{1}{2}(1-\frac{\eta}{\xi}) & \text{for } 0<\eta
        \end{cases}
    \end{equation}

\end{appendix}

\clearpage
\bibliography{esqpt}

\end{document}